\documentclass[conference]{IEEEtran}
\IEEEoverridecommandlockouts
\usepackage{cite}
\usepackage[numbers,sort&compress]{natbib}
\usepackage{amsmath,amssymb,amsfonts}
\usepackage{algorithmic}
\usepackage{graphicx}
\usepackage{textcomp}
\usepackage{xcolor}
\usepackage{subcaption}
\usepackage{booktabs}
\def\BibTeX{{\rm B\kern-.05em{\sc i\kern-.025em b}\kern-.08em
    T\kern-.1667em\lower.7ex\hbox{E}\kern-.125emX}}

\begin{document}
\title{Enhancing Multimodal Emotion Recognition through Multi-Granularity Cross-Modal Alignment
\thanks{\dag Corresponding author. This work was supported by the National Key R$\&$D Program of China (Grant No.2022ZD0116307) and NSF China (Grant No.62271270).}
}

\author{\IEEEauthorblockN{Xuechen Wang\IEEEauthorrefmark{1}, Shiwan Zhao\IEEEauthorrefmark{1}, Haoqin Sun\IEEEauthorrefmark{1}, Hui Wang\IEEEauthorrefmark{1}, Jiaming Zhou\IEEEauthorrefmark{1}, and Yong Qin\IEEEauthorrefmark{1}$^{\dag}$}
\IEEEauthorblockA{\IEEEauthorrefmark{1}TMCC, College of Computer Science, Nankai Unversity, Tianjin, China\\
Email: shirleywxc0103@mail.nankai.edu.cn}
}

\maketitle


\begin{abstract}
Multimodal emotion recognition (MER), leveraging speech and text, has emerged as a pivotal domain within human-computer interaction, demanding sophisticated methods for effective multimodal integration. The challenge of aligning features across these modalities is significant, with most existing approaches adopting a singular alignment strategy. Such a narrow focus not only limits model performance but also fails to address the complexity and ambiguity inherent in emotional expressions. In response, this paper introduces a \textbf{Multi-Granularity Cross-Modal Alignment (MGCMA)} framework, distinguished by its comprehensive approach encompassing distribution-based, instance-based, and token-based alignment modules. This framework enables a multi-level perception of emotional information across modalities. Our experiments on IEMOCAP demonstrate that our proposed method outperforms current state-of-the-art techniques.
\end{abstract}

\begin{IEEEkeywords}
multimodal emotion recognition, multi-granularity alignment, human-computer interaction
\end{IEEEkeywords}

\section{Introduction}
Emotion recognition is an important aspect of human-computer interaction. Extensive research has been conducted on unimodal emotion recognition \cite{adoma2020comparative, wang23q_interspeech}. However, emotions are conveyed in a variety of ways. Speech and text are both important carriers of emotional information \cite{peng2021efficient}. Multimodal emotion recognition (MER), integrating speech and text, has gained significant attention \cite{liu2022discriminative}. By fusing information from multiple modalities, such as prosodic features from speech and semantic cues from text, MER aims to capture a more comprehensive understanding of human emotions. In recent years, MER has been widely used in many applications across various domains \cite{hu2018touch, arsikere2014computationally}. 

In the field of MER, the heterogeneity across modalities makes the alignment of multimodal features become a hot research topic. 
Effective alignment methods are crucial for leveraging complementary information from different modalities. 
Sun et al. \cite{sun2023fine} employ a fine-grained alignment component to obtain modality-shared representations and capture modal consistency. Liu et al. \cite{liu2021multimodal} integrate the alignment module with a silence removal technology, achieving accurate alignment of speech and text. These approaches have yielded favorable results in MER. However, they have not fully addressed the ambiguity inherent in emotional expressions \cite{zhou2022multi}, which can compromise the quality of alignment. Drawing inspiration from \cite{Ji_2023_CVPR}, leveraging distribution-level representation as a higher-dimensional tool for coarse-grained alignment offers a promising approach to model such ambiguity.

On the other hand, to achieve fine-grained alignment and uncover potential mapping relationships between multimodal inputs, various innovative alignment strategies have been proposed. Among these, the cross-modal attention mechanism stands out, having been extensively applied to facilitate token-level alignment in numerous studies \cite{choi2018convolutional, n20_interspeech}. For instance, Liu et al. \cite{liu20b_interspeech} devise an attention-based bidirectional alignment network to map the alignment between speech and text. To address the challenges of emotional asynchrony and modality misalignment in MER, Fan et al. \cite{fan2023mgat} introduce a novel multi-granularity attention mechanism to enhance alignment accuracy. However, token-level alignment predominantly concentrates on local information, frequently overlooking the global context. This oversight can, to some extent, compromise the model's overall performance when token-level alignment is applied in isolation.

\begin{figure*}[t]
  \centering
  \includegraphics[width=0.9\linewidth]{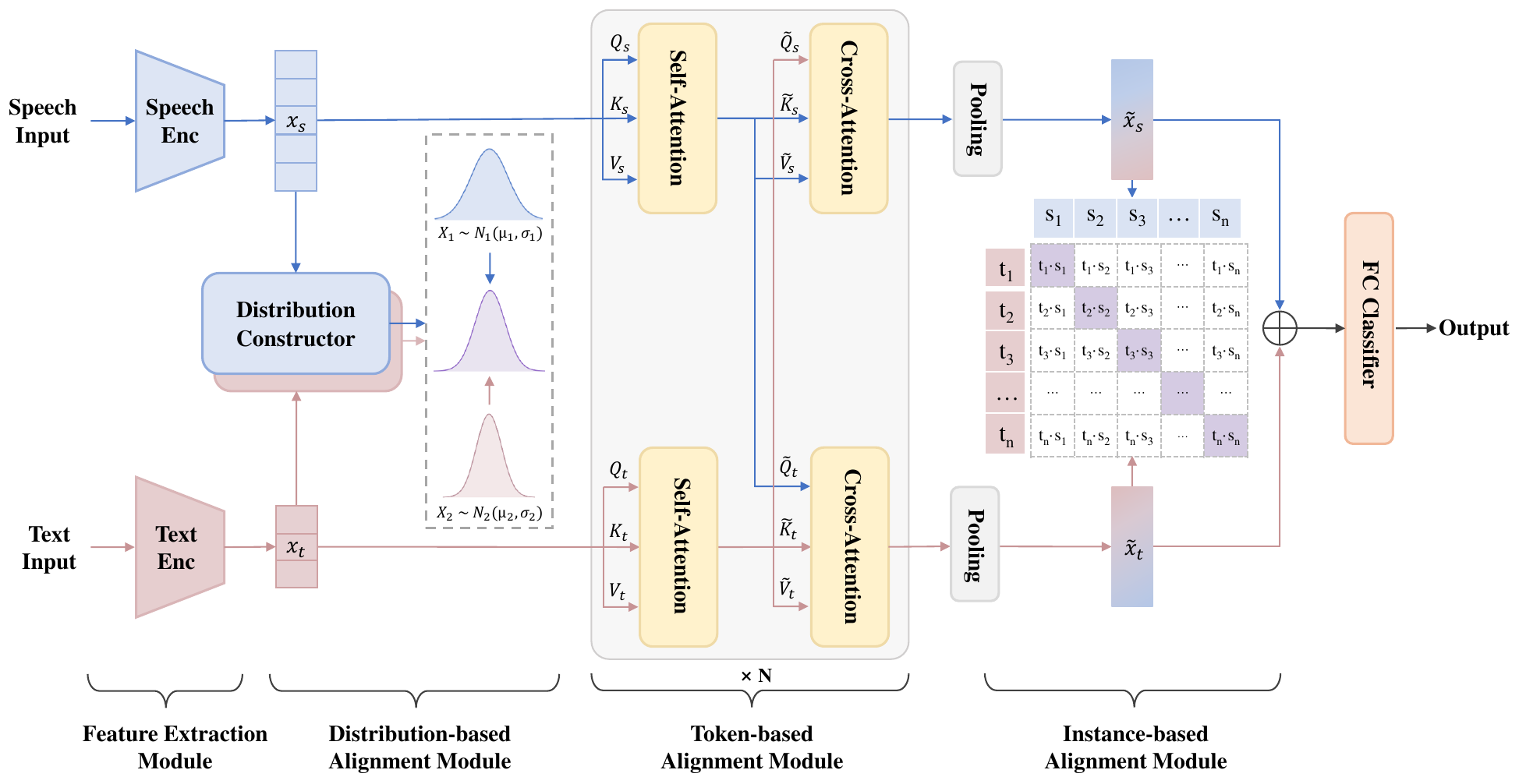}
\caption{Overview of our proposed Multi-Granularity Cross-Modal Alignment (MGCMA) framework which comprises distribution-based, token-based, instance-based alignment modules, and a feature extractor.}
  \label{fig:overview}
\end{figure*}

Conversely, contrastive learning has emerged as an innovative method for achieving instance-level multimodal feature alignment \cite{radford2021learning, elizalde2023clap}. This approach, by maximizing the similarity between positive pairs and minimizing it for negative pairs, enables the model to develop more generalized representations. Empirical evidence underscores the effectiveness of contrastive learning in enhancing multimodal alignment. Nonetheless, the scope of existing research is often limited to single-level alignment—either at the token or instance level—thereby restricting the model’s ability to fully grasp emotional information across different levels.

In this paper, to address the aforementioned challenges in MER, we introduce a novel \textbf{Multi-Granularity Cross-Modal Alignment (MGCMA)} framework which comprises distribution-based, token-based, and instance-based alignment modules. The distribution-based alignment module, implemented through distribution-level contrastive learning, acts as a 
coarse-grained alignment from a higher-dimensional perspective, accommodating the ambiguity of emotions. By aligning distribution representations that encapsulate richer information, it sets the stage for more nuanced alignments and interactions. The token-based alignment module employs self-attention and cross-attention mechanisms to accurately match and align local cross-modal representations between speech and text. This fine-grained alignment facilitates the exchange of information across modalities, which is advantageous for the subsequent instance-level alignment. 
The instance-based alignment module prompts the model to learn improved mapping relationships and bolster the correlation between specific speech-text pairs. Our multi-granularity cross-modal alignment framework enables the model to achieve a multi-level perception of emotional information. It integrates the strengths of various granularity alignment strategies and addresses the previously overlooked issue of ambiguity in emotional expression.

In experiments conducted on the IEMOCAP dataset, our proposed framework surpasses current state-of-the-art methods, achieving weighted accuracy (WA) of 78.87\% and unweighted accuracy (UA) of 80.24\%.



\vspace{-1pt}
\section{Proposed Method}
As shown in Fig. \ref{fig:overview}, our proposed method consists of a feature extractor, a distribution-based alignment module, a token-based alignment module, and an instance-based alignment module. The following subsections detail each of these components.

\vspace{-2pt}
\subsection{Feature Extractor}
To obtain high-level representations of each modality, we utilize self-supervised pre-trained models Wav2vec2.0 \cite{baevski2020wav2vec} and BERT \cite{devlin-etal-2019-bert} as our speech and text encoders, respectively. For each speech-text pair, the speech input $s$ is encoded into $x_{s}$ $\in$ $R^{L_s \times D_s}$, and the text input $t$ is encoded into $x_{t}$ $\in$ $R^{L_t \times D_t}$, where \textit{L} represents the sequence length and \textit{D} represents the embedding dimension.

\vspace{-2pt}
\subsection{Distribution-based Alignment Module}
\subsubsection{Distribution Constructor}
\begin{figure}[t]
  \centering
  \includegraphics[width=0.82\linewidth]{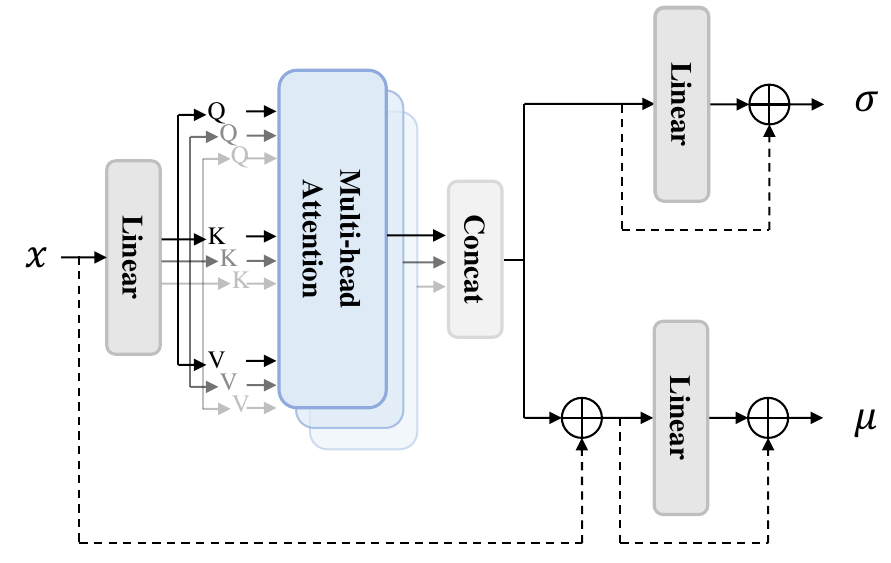}
  \caption{The structure of Distribution Constructor. Activation layers and normalization layers are omitted in the diagram.}
  \label{fig:Distribution}
  \vspace{-8pt}
\end{figure}

In order to overcome the shortcomings of the traditional representations, we apply a distribution constructor to construct a multivariate Gaussian distribution for the features $x_{s}$ and $x_{t}$. As shown in Fig. \ref{fig:Distribution}, our distribution constructor is implemented by the multi-head self-attention mechanism. 

Given a conventional feature representation $x_m$ $\in$ $R^{L_m \times D_m}$, \textit{m $\in$ \{s, t\}}, the linear layer projects it to \textit{Q, K, V} and splits it into \textit{k} heads: $Q_i$, $K_i$ and $V_i$, \textit{i = \{1, …, k\}}. The multi-head self-attention is performed as follows:
\begin{align}
\operatorname{Attention}(Q,K,V)=\operatorname{softmax}\left(\frac{{Q}{K}^T}{\sqrt{d}}\right){V},\\head_i=\text{ Attention }\left({Q_i}W_i^q,{K_i}W_i^k,{V_i}W_i^v\right),\\\operatorname{Concat}=\left[head_1, head_2, \ldots, head_k\right]W^o,
\end{align}
where $W_i^q$, $W_i^k$, $W_i^v$ and $W^o$ are trainable matrices, $d$ is set to $D/k$. The $Concat$ result is divided into $\mu$ and ${\sigma}$ branches. Linear layers and residual connections are applied to obtain the final output. 

\subsubsection{Distribution-level Contrastive Learning}
After constructing distributions of features from two modalities, we implement distribution-based alignment module by conducting a distribution-level contrastive learning. Given two multivariate Gaussian distributions $\mathcal{N}_1\sim(\mu_1,\Sigma_1)$ and $\mathcal{N}_2\sim(\mu_2,\Sigma_2)$, we first calculate the 2-Wasserstein distance between them as follows:
\begin{align}
\nonumber W(\mathcal{N}_1, \mathcal{N}_2) &= ||\mu_1-\mu_2||_2^2 \\
\nonumber&\quad+ \mathrm{Tr}(\Sigma_1+\Sigma_2-2(\Sigma_1^{1/2}\Sigma_2\Sigma_1^{1/2})^{1/2})\\
\nonumber&= ||\mu_1-\mu_2||_2^2 + \mathrm{Tr}((\Sigma_1^{1/2}-\Sigma_2^{1/2})^2)\\
&= ||\mu_1-\mu_2||_2^2 + ||\sigma_1-\sigma_2||_2^2,
\end{align}
where $\Sigma_1^{1/2}\Sigma_2\Sigma_1^{1/2}$ could be written as $\Sigma_1\Sigma_2$ because $\Sigma_1$ and $\Sigma_2$ are both diagonal matrices. Based on the above distances, we calculate the similarity of two distributions as follows:
\begin{align}
Sim(\mathcal{N}_1, \mathcal{N}_2)=-p\cdot W(\mathcal{N}_1, \mathcal{N}_2)+q,
\end{align}
where \textit{p (p$\textgreater$0)} denotes a scaler factor and \textit{q} denotes the bias. This formula reflects a negative correlation between the distance and similarity of distributions.

For a set of \textit{2N} instances containing \textit{N} speech-text pairs \textit{\{$s_i$, $t_i$\}, i = 1, …, N}, we obtain \textit{2N} distributions \textit{$\mathcal{N}_{s_i}\sim(\mu_{s_i},\Sigma_{s_i})$} and \textit{$\mathcal{N}_{t_i}\sim(\mu_{t_i},\Sigma_{t_i})$}. Instances from the same speech-text pair are treated as \textit{positives}, while the others are regarded as \textit{negatives}. The distribution-level contrastive learning loss and the final distribution-based alignment loss $\mathcal{L}_{\mathrm{DA}}$ are calculated as follows:
\begin{align}
\mathcal{L}^{s2t}_i&=-\log\frac{\exp(Sim(\mathcal{N}_{s_i}, \mathcal{N}_{t_i})/\tau)}{\sum_{n=1}^N\exp(Sim(\mathcal{N}_{s_i}, \mathcal{N}_{t_n})/\tau)},\\
\mathcal{L}^{t2s}_i&=-\log\frac{\exp(Sim(\mathcal{N}_{t_i}, \mathcal{N}_{s_i})/\tau)}{\sum_{n=1}^N\exp(Sim(\mathcal{N}_{t_i}, \mathcal{N}_{s_n})/\tau)},\\
\mathcal{L}_{DA}&=\frac1{2N}\sum_{i=1}^{N}(\mathcal{L}^{s2t}_i+\mathcal{L}^{t2s}_i),
\end{align}
where $\tau$ denotes a scalar temperature parameter, \textit{N} is the total number of speech-text pairs, $\mathcal{L}^{s2t}_i$ and $\mathcal{L}^{t2s}_i$ denotes the speech-to-text contrastive loss and text-to-speech contrastive loss of the $\textit{i}^{th}$ input \textit{\{$s_i$, $t_i$\}}, respectively. 

\subsection{Token-based Alignment Module}
The distribution-based alignment module facilitates more effective alignment of different components within an utterance. Subsequently, we employ the token-based alignment module to achieve fine-grained alignment and foster extensive interaction between features from different modalities. This token-based alignment module comprises \textit{N} blocks, each consisting of self-attention and cross-attention mechanisms. For the \textit{i}-th block, the inputs for self-attention, denoted as ${Q}, {K}, {V}$, are derived from the \textit{(i-1)}-th block's cross-attention outputs, while the inputs for cross-attention, denoted as ${\widetilde{Q}}, {\widetilde{K}}, {\widetilde{V}}$, are sourced from the \textit{i}-th block's self-attention outputs. Consequently, this iterative process yields text-aware speech representations ${\widetilde{x}}_{s}$ and speech-aware text representations ${\widetilde{x}}_{t}$.


\vspace{-2pt}
\subsection{Instance-based Alignment Module}
To further enhance the association of the local information and achieve better mappings between specific speech-text pairs, we implement an instance-based alignment. It is implemented by a contrastive learning with the representations ${\widetilde{x}}_{s}$ and ${\widetilde{x}}_{t}$. The instance-level contrastive learning loss and the final instance-based alignment loss $\mathcal{L}_{\mathrm{IA}}$ are calculated as follows:
\vspace{-2pt}
\begin{align}
\mathcal{\widetilde{L}}^{s2t}_i&=-\log\frac{\exp(\boldsymbol{\widetilde{x}}_{s_i}\cdot\boldsymbol{\widetilde{x}}_{t_i})/\tau)}{\sum_{n=1}^N\exp(\boldsymbol{\widetilde{x}}_{s_i}\cdot\boldsymbol{\widetilde{x}}_{t_n})/\tau)},\\
\mathcal{\widetilde{L}}^{t2s}_i&=-\log\frac{\exp(\boldsymbol{\widetilde{x}}_{t_i}\cdot\boldsymbol{\widetilde{x}}_{s_i})/\tau)}{\sum_{n=1}^N\exp(\boldsymbol{\widetilde{x}}_{t_i}\cdot\boldsymbol{\widetilde{x}}_{s_n})/\tau)},\\
\mathcal{L}_{IA}&=\frac1{2N}\sum_{i=1}^{N}(\mathcal{\widetilde{L}}^{s2t}_i+\mathcal{\widetilde{L}}^{t2s}_i),
\end{align}
\vspace{-1pt}
where ($\cdot$) symbol denotes the dot product, $\mathcal{\widetilde{L}}^{s2t}_i$ and $\mathcal{\widetilde{L}}^{t2s}_i$ denotes the speech-to-text contrastive loss and text-to-speech contrastive loss of the $\textit{i}^{th}$ input \textit{\{$s_i$, $t_i$\}}. $\tau$ and \textit{N} have the same meaning as before. 

Building upon the above alignment stategies, we additionally incorporate cross-entropy loss $\mathcal{L}_{CE}$ for supervised training at the end. The final learning objective is expressed as follows:
\vspace{-1pt}
\begin{align}
\mathcal{L}=\mathcal{L}_{DA}+\mathcal{L}_{IA}+\mathcal{L}_{CE}.
\end{align}

\vspace{-1pt}
\section{Experiments}

\vspace{-1pt}
\subsection{Dataset}
We assess the effectiveness of our proposed method using the IEMOCAP \cite{busso2008iemocap} dataset. To align with previous studies, we perform experiments on a subset of four emotions: angry, happy, sad, and neutral. We combine the original happy category and excited category into a single happy category. In total, the dataset contains 5,531 utterances, which includes 1,103 angry, 1,636 happy, 1,084 sad, and 1,708 neutral utterances.

\vspace{-1pt}
\subsection{Experimental Settings}
In this work, we use the publicly available pre-trained models, \textit{Wav2Vec 2.0 Base} and \textit{Bert-base-uncased}. The dimension size of feature representations is set to 768. The head number of attention mechanism is set to 12 and the block number of token-based alignment module is set to 6. We apply 5-fold cross-validation to evaluate our results, leaving one session out for testing. The batch size is set to 4 and the max training epoch is set to 100. We choose Adam optimizer with the initial learning rate of 10$^{-5}$. The hyperparameter $\tau$ is set to 0.07. We use weighted accuracy (WA) and unweighted accuracy (UA) as metrics to evaluate the performance of our proposed method.

\subsection{Comparison with SOTA Approaches}
Recent state-of-the-art (SOTA) approaches are presented in Table \ref{tab:sota}. Compared with these results, our method MGCMA shows a significant improvement on both WA and UA metrics. The outstanding performance can be attributed to the model's multi-level perceptual ability towards emotional information, which demonstrates the effectiveness of our proposed method.

\begin{table}
  \caption{Performance comparison of our proposed method with SOTA approaches on IEMOCAP.}
  \label{tab:sota}
  \centering
  \begin{tabular}{llll}
    \toprule
     \textbf{Method} & \textbf{Year} & \textbf{WA(\%)} & \textbf{UA(\%)}\\
    \midrule
    Chen et al. \cite{chen2022key} & 2022 & 74.30 & 75.30\\
    Sun et al. \cite{sun2023using} & 2023 & 78.42 & 79.71\\
    Zhao et al. \cite{zhao2023knowledge} & 2023 & 75.50 & 77.00\\
    Wang et al. \cite{wang2023exploring} & 2023 & 75.20 & 76.40\\
    Zhang et al. \cite{zhang23g_interspeech} & 2023 & 76.00 & 77.80\\
    Zhao et al. \cite{zhao23b_interspeech} & 2023 & 77.40 & 78.50\\
    \midrule
    \textbf{MGCMA (ours)} & \textbf{2024} & \textbf{78.87} & \textbf{80.24}\\
    \bottomrule
  \end{tabular}
  \vspace{-9pt}
\end{table}

\subsection{Ablation Studies and Analysis}
We conduct a series of experiments to demonstrate the effectiveness and necessity of each module in our framework. The results are shown in Table \ref{tab:all results}, where DAM, TAM and IAM denotes the distribution-based, token-based and instance-based alignment module, respectively. 
More detailed analysis are discussed below.

\begin{table}
\caption{Results of ablation studies on IEMOCAP.}
  \label{tab:all results}
  \centering
  \begin{tabular}{llll}
    \toprule
    \textbf{System} & \textbf{Method} & \textbf{WA(\%)} & \textbf{UA(\%)}\\
    \midrule
    \textbf{S0} & \textbf{MGCMA}  & \textbf{78.87} & \textbf{80.24}\\
    S1 & w/o DAM & 77.78 & 79.07\\
    S2 & w/o TAM & 78.00 & 79.14\\
    S3 & w/o IAM & 78.30 & 79.38\\
    S4 & w/o (DAM + TAM + IAM) & 76.20 & 77.62\\
    \bottomrule
  \end{tabular}
  \vspace{-6pt}
\end{table}

\textbf{Importance of Different Granularity Alignment Modules.}
Upon examining the results in Table \ref{tab:all results}, it becomes apparent that DAM exerts the most significant influence on the overall framework. Within a multimodal framework, alignment at a high-dimensional level enables the model to facilitate more effective token-based and instance-based alignments subsequently.

The effect of TAM is demonstrated by comparing the results of S0 and S2. Incorporating TAM allows features from one modality to assimilate information from the other modality, thereby augmenting the interaction among local features pertinent to emotion.

A comparison between the results of S0 and S3 highlights the efficacy of IAM. This module compels the framework to position matched speech-text pairs closer, while distancing unmatched pairs. With better distributions in the latent space, the model can obtain a stronger recognition capacity.

\begin{figure}[t]
	\centering
	\begin{subfigure}{0.47\linewidth}
		\centering
		\includegraphics[width=1\linewidth]{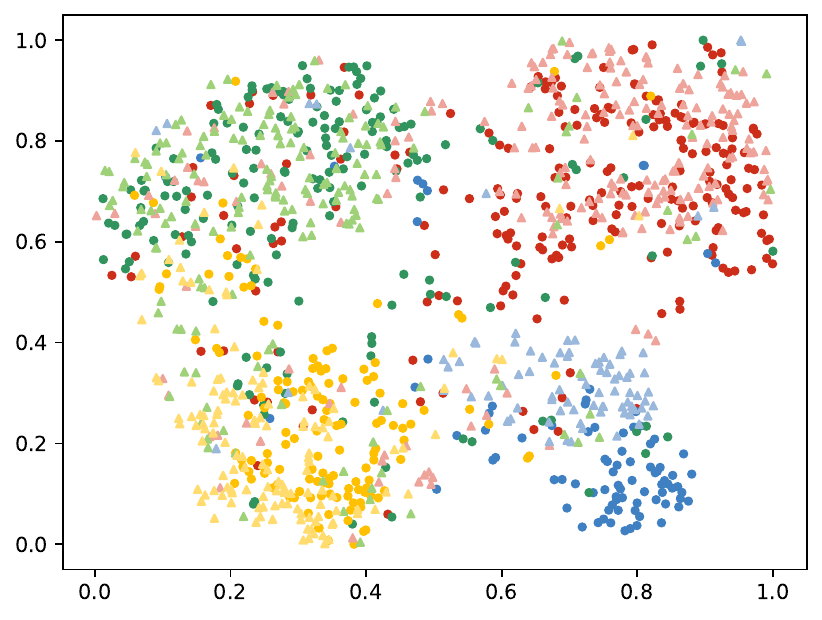}
		\caption{MGCMA}
		\label{MGCMA_tsne}
	\end{subfigure}
	\centering
	\begin{subfigure}{0.47\linewidth}
		\centering
		\includegraphics[width=1\linewidth]{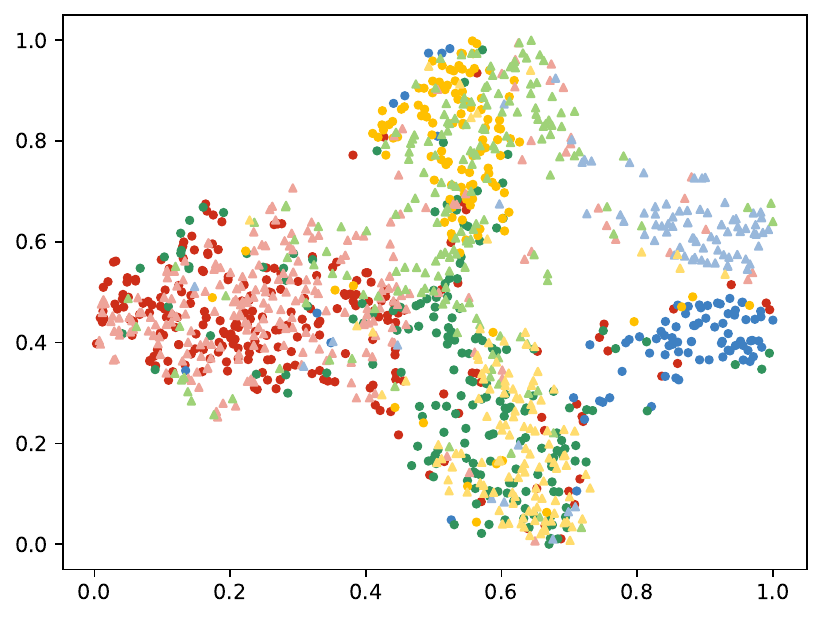}
		\caption{w/o DAM}
		\label{without_dam_tsne}
	\end{subfigure}\\
	\centering
	\begin{subfigure}{0.47\linewidth}
		\centering
		\includegraphics[width=1\linewidth]{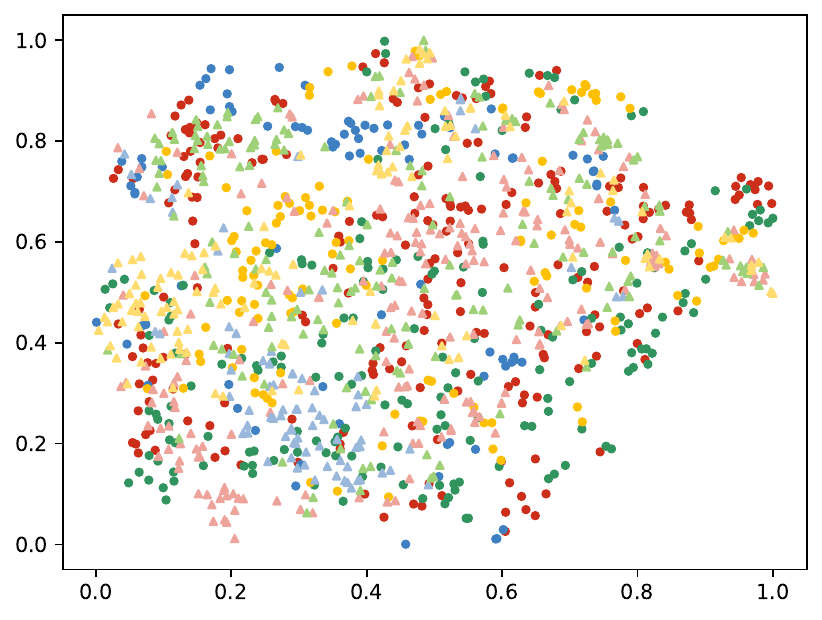}
		\caption{w/o TAM}
		\label{without_tam_tsne}
	\end{subfigure}
        \begin{subfigure}{0.47\linewidth}	
        \centering
		\includegraphics[width=1\linewidth]{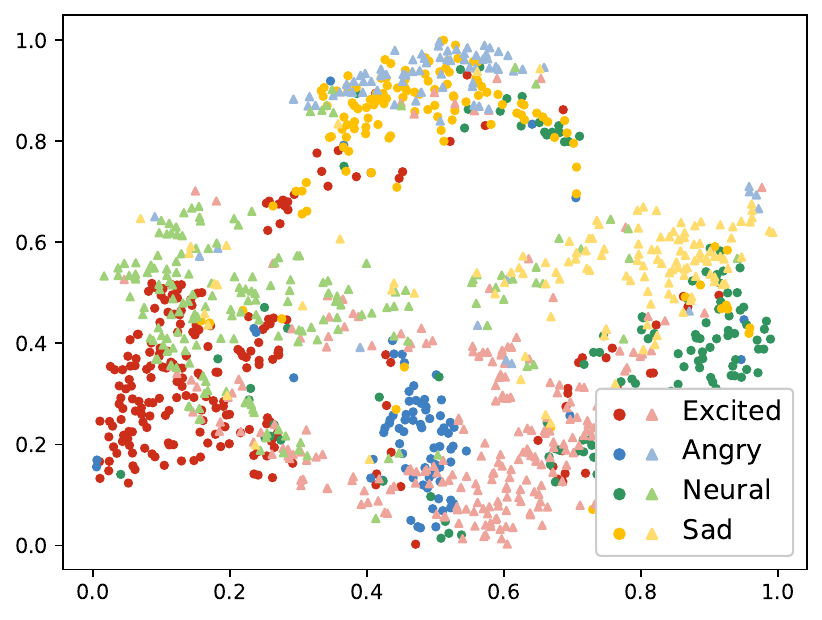}
		\caption{w/o IAM}
		\label{without_iam_tsne}
	\end{subfigure}
   	\caption{Visualization of the representations with different alignment strategies, where darker colors indicate speech features and lighter colors indicate text features. 
}
	\label{TSNE}
 \vspace{-6pt}
\end{figure}

To intuitively illustrate the distribution of speech and text features under various alignment strategies, we employ the t-SNE technique \cite{2008Visualizing} to visualize the outputs of the intermediate layer. As shown in Fig \ref{TSNE}, our proposed MGCMA achieves the most effective alignment results.



\textbf{Importance of the sequence of different alignment modules.}
To further investigate the impact of varying sequences of alignment modules on the results, we shuffle the order of these modules. The performances are documented in Table \ref{tab:sequences}.

It can be observed that the S0 alignment sequence yields the most favorable outcomes. Serving as a higher-dimensional feature, distribution representation contains more information than conventional methods. The distribution-based alignment module enhances the consistency of various segments in an utterance, playing a pivotal role in elevating the performance of token-based alignment. Through the token-based alignment module, the model focuses more on local information pertinent to the other modality, extracting features critical for emotion recognition. 
Subsequently, the instance-based alignment module strengthens the association of multimodal features and fosters accurate correlation between the correct speech-text pairs, which facilitates the instance-based downstream task. 
Altering the sequence of the alignment modules disrupts the model's ability to develop a multi-level perception of emotional information, which can be detrimental to performance.

\begin{table}
  \caption{Performances with different sequences of alignment modules. }
  \label{tab:sequences}
  \centering
  \begin{tabular}{llll}
    \toprule
    \textbf{System} & \textbf{Sequence} & \textbf{WA(\%)} & \textbf{UA(\%)}\\
    \midrule
    \textbf{S0} & \textbf{DAM + TAM + IAM}  & \textbf{78.87} & \textbf{80.24}\\
    S5 & DAM + IAM + TAM & 78.76 & 79.48\\
    S6 & IAM + DAM + TAM & 77.60 & 78.60\\
    S7 & IAM + TAM + DAM & 78.71 & 79.81\\
    S8 & TAM + DAM + IAM & 77.33 & 78.75\\
    S9 & TAM + IAM + DAM & 76.88 & 78.43\\
    \bottomrule
  \end{tabular}
  \vspace{-9pt}
\end{table}

\vspace{-4pt}
\section{Conclusions}
In this paper, we propose a novel Multi-Granularity Cross-Modal Alignment (MGCMA) framework for multimodal emotion recognition. Our approach encourages the model to exhibit comprehensive processing capabilities for emotional information across different granularities. We conduct a series of experiments to prove the effectiveness and necessity of our proposed method. In comparison to state-of-the-art results, our proposed method shows significant improvements on both WA and UA.






\end{document}